\documentclass[aps,prc,groupedaddress,twocolumn]{revtex4-1}

\usepackage{amsmath} 
\usepackage{amssymb} 
\usepackage{graphicx} 

\begin{document}

\title{Second random-phase approximation, Thouless' theorem and the stability condition reexamined and clarified} 

\author{P. Papakonstantinou}
\email{ppapakon@ibs.re.kr} 
\affiliation{Rare Isotope Science Project, Institute for Basic Science, Daejeon 305-811, Republic of Korea}

\date{\today}

\begin{abstract}
It has been revealed through numerical calculations that the Second Random Phase Approximation (SRPA) 
with the Hartree-Fock solution as its reference state 
results in 1) spurious states at genuinely finite energy, contrary to common expectation, and 
2) unstable solutions, which within the first-order Random Phase Approximation correspond to real low-energy collective vibrations. 
In the present work, these shortcomings of SRPA 
are shown to not contradict Thouless' theorem about the energy-weighted sum rule,  
and their origin is traced to the violation of the stability condition. 
A more general theorem is proven. 
Formal arguments are elucidated through numerical examples. 
Implications for the validity of SRPA are discussed. 
\end{abstract}


\maketitle

\section{Introduction}


In nuclear theory, the random-phase approximation (RPA)~\cite{Row1970,RS80} has proven a valuable tool in the description of  
collective nuclear vibrations, foremost giant resonances. 
In its self-consistent version, 
it facilitates connections between the properties of such vibrations and 
the equation of state of nuclear matter. 
The self-consistent RPA for closed-shell nuclei is understood here as the formulation of RPA whereby the reference state is 
the solution of the Hartree-Fock (HF) equations for a given nuclear interaction or density functional, 
while the RPA residual interaction is derived from the same interaction or functional. 
We may denote this method by ``HF-RPA". 

It can be argued that HF-RPA is not a self-consistent many-body method, in the sense that the HF solution differs from the correlated RPA ground state. 
As long as  we are interested in (long-range) density fluctuations, however, and the HF ground state describes well the long-range spatial properties of nuclei, e.g., the root mean square radius, the use of a correlated ground state in first-order RPA makes only a marginal difference~\cite{PRP2007}. 
This outcome validates the equations-of-motion philosophy~\cite{Row1968a} of working with low-rank (in the many-body sense) operators. The latter are obtained as (double) commutators of higher-rank operators and the sensitivity of their matrix elements to correlations in the ground state is reduced by comparison. 

HF-RPA is self-consistent in other important ways, though. 
The HF and RPA methods are intimately connected~\cite{Tho1960,Tho1961}: 
The HF equations, which constitute a variational problem, are equivalent to the stability condition of RPA~\cite{Tho1960}. 
If HF-RPA has imaginary solutions, these must signify a phase transition. 
For example, an imaginary $J^{\pi}=2^+$ solution to HF-RPA in a spherical basis indicates that the nuclear ground state for the given Hamiltonian cannot be spherical. 
In the absence of such instabilities, 
the summed energy-weighted strength of a single-particle transition operator in RPA is determined by a specific expectation value  
in the HF ground state. 
Symmetries which are broken by the HF ground state, such as translational invariance in the case of a localized nuclear wave-function, are restored by HF-RPA: spurious ``excitations" such as center-of-mass translations, are guaranteed to appear at zero excitation energy, according to Thouless' theorem~\cite{Tho1961}. 
Thanks to the above formal and important properties, 
the HF-RPA method deserves the attribute ``self-consistent". 

Among extensions of RPA to include more complex configurations we find the Second (or second-order) RPA (SRPA)~\cite{daP1965,Yan1987}, where the vibration operators are constructed not only with particle-hole ($ph$), but also with two-particle-two-hole ($2p2h$) configurations. 
SRPA accommodates  collisional damping and has been used to describe the fragmentation width of resonances~\cite{Wam1988,DNS1990,LAC2004,Usm2011,GGC2011}. 
Applications in metallic clusters exist as well~\cite{GaC2009}. 
Recent formal progress includes the derivation of rearrangement terms to be taken into account when employing 
density-dependent forces~\cite{GGC2011a}. 
Early applications in nuclear physics~\cite{Wam1988,DNS1990} were realized in restricted $2p2h$ spaces, comprising configurations energetically near the resonances of interest. 
The single-particle spectrum was typically that of a Woods-Saxon potential.  
Often the ground state included correlations. 
Ground-state correlations are particularly relevant for the description of Gammow-Teller and other low-energy resonances. 
Formalisms going beyond RPA in the spirit of SRPA are known as extended-RPA theories and are often formulated on a correlated ground state~\cite{DNS1990,TSA1988,AdL1988,ToS2004,GGC2006,ToS2007}.

In analogy to HF-RPA, one may speak of a ``self-consistent" HF-SRPA, where the ground state is the HF solution. 
Formal properties analogous to HF-SRPA~\cite{Yan1987} have made the method particularly alluring 
and led to several recent applications~\cite{PaR2009,Usm2011,GGC2011,GGD2012}. 
The first fully self-consistent HF-SRPA calculations, where all couplings in the HF solutions and the SRPA matrix are described by the same Hamiltonian, and the full $ph$ and $2p2h$ basis generated by the HF solution is used, were reported in \cite{PaR2009,PaR2010}. 
The promise, as it were, of HF-SRPA was to provide a self-consistent framework for the description of a variety of nuclear vibrations using realistic, renormalized finite-range potentials, not fitted to HF or RPA results, and therefore not causing double-counting.  
Besides increased computing power and new solution techniques,  
the calculations were made possible thanks to a two-body, finite-range potential, which does not cause divergencies. 
By contrast, applications with zero-range Skyrme interactions must 
use a truncated $2p2h$ space; not only to minimize double counting of higher-order effects (which is unavoidable), but also because there is no natural momentum cutoff. 

The above numerical applications~\cite{PaR2009,PaR2010} demostrated the formal properties of HF-SRPA for the first time: the precise conservation of the energy-weighted sum rule despite a strong downward energetic shift of the dominant resonances (see also Refs.~\cite{GaC2009,GGD2012}); and the generation of self-energy corrections and fragmentation. 
Unfortunately, they also revealed that the HF-SRPA method  
results in spurious states at genuinely finite energy, and 
in unstable solutions, which within HF-RPA correspond to real low-energy collective vibrations. 
The above are in apparent contrast with the formal derivations of Ref.~\cite{Yan1987}, which follow from Thouless' theorem~\cite{Tho1961}. 

It is the purpose of the present article to clarify the situation. 
It will be shown, with the help of illuminating numerical examples, that the problems encountered in Refs.~\cite{PaR2009,PaR2010} are 
inherent in the HF-SRPA formalism; that they do not contradict Thouless' theorem; and that the contradiction with Ref.~\cite{Yan1987} is caused by the violation of the stability condition.

In what follows, by ``SRPA" we will mean ``HF-SRPA" as a rule--similarly for RPA. 
 
It is important to first be aware of what has {\em not} been proven about SRPA. 
As a result of not {\em proving} much yet, some arguments in Sec.~\ref{Sec:Formal} and Sec.\ref{Sec:Violation} may appear speculative. 
Their relevance will become clear through the rest of Sec.~\ref{Sec:Stability}, where the stability condition is addressed and Thouless' theorem derived and interpreted anew, and in Sec.~\ref{Sec:Numerical}, where numerical examples are analyzed. 
We provide an outlook in Sec.~\ref{Sec:Summary}.

\section{Formal observations \label{Sec:Formal}} 

We consider the SRPA as it was formulated in, e.g., Ref.~\cite{Yan1987}. 
In brief, excited states $|\nu\rangle$ of energy $E_{\nu}$ with respect to the ground state are considered as combinations of 
$ph$ and $2p2h$ configurations, 
\begin{align} 
|\nu\rangle = &  
\sum_{ph} (X^{(1)}_{\nu; ph} a^{\dagger}_pa_h 
-Y^{(1)}_{\nu;ph} a^{\dagger}_ha_p )  
\nonumber \\ 
 & +  \sum_{p_1p_2h_1h_2} (X^{(2)}_{\nu ; p_1p_2h_1h_2} a^{\dagger}_{p_1}a^{\dagger}_{p_2} a_{h_2}a_{h_1} 
\nonumber \\ 
 & -  Y^{(2)}_{\nu ; p_1p_2h_1h_2} a^{\dagger}_{h_1}a^{\dagger}_{h_2} a_{p_2}a_{p_1} 
) . 
\end{align} 
The HF ground state is considered as the reference state. 
The arrays containing the forward 
$X_{\nu}^{\dagger}=(X^{(1)\dagger}_{\nu} X^{(2)\dagger}_{\nu})$ 
and backward 
$Y_{\nu}^{\dagger}=(Y^{(1)\dagger}_{\nu} Y^{(2)\dagger}_{\nu})$ 
amplitudes are the solutions of the SRPA equations in the $ph+2p2h$ space,
\begin{equation} 
\left( 
\begin{array}{cc} 
A & B \\
-B^{\ast} & -A^{\ast} 
\end{array} 
\right) 
\left( 
\begin{array}{c}  
X_{\nu} \\ Y_{\nu} 
\end{array} 
\right) 
= 
E_{\nu} 
\left( 
\begin{array}{c}  
X_{\nu} \\ Y_{\nu} 
\end{array} 
\right) 
\, . 
\label{Eq:SRPA} 
\end{equation}  
The matrices $A$ (Hermitean) and $B$ (symmetric) are block matrices in the $ph+2p2h$ space, 
\[ 
A = 
\left( 
\begin{array}{cc} 
A_{11} & A_{12} \\ A_{21} & A_{22} 
\end{array} 
\right) 
\,\, ; \,\,  
B = 
\left( 
\begin{array}{cc} 
B_{11} & B_{12} \\ B_{21} & B_{22} 
\end{array} 
\right) 
\, . 
\] 
In the above, $A_{11}$ and $B_{11}$ are the usual RPA matrices in the $ph$ space, 
the subscripts ``$12$" or ``$21$"  denote the matrices coupling the $ph$ and the $2p2h$ spaces and the subscripts ``$22$" denote matrices in the $2p2h$ space. 
The matrix elements are defined as double commutators of the nuclear Hamiltonian $H$ with the vibration creation 
and annihilation operators~\cite{Row1968a,TSA1988,AdL1988}.  
For two-body Hamiltonians (and a HF reference state)  we have 
\[ B_{12} = B_{22} = 0 . \] 
The matrices $A$ and $B$ are of dimension $M\times M$, where $M=M_{ph}+M_{2p2h}$ in an obvious notation. 
The usual Tamm-Dancoff approximations to RPA and SRPA (TDA and STDA, respectively) are obtained by setting $B=0$. 
They are Hermitean eigenvalue problems of dimension $M_{ph}\times M_{ph}$ (first-order) or $M\times M$ (second-order) 
 and with real eigenvalues. 

The $2M\times 2M$ (S)RPA eigenvalue problem (\ref{Eq:SRPA})  
can be reduced to a problem of dimension $M\times M$~\cite{UlG1972,Pap2007}. 
The derivations of Ref.~\cite{Pap2007} can be used to analyze the properties of the SRPA matrix in terms of the $A\pm B$ submatrices, when those are real 
(for complex matrices see Ref.~\cite{UlG1972}).
In particular, we note that: 
i) If both $A\pm B$ are positive-definite, all (S)RPA solutions are real; ii) if one of the two 
matrices $A\pm B$ has one negative eigenvalue, we should expect a pair of imaginary (S)RPA eigenvalues; 
iii) if both $A\pm B$ have one negative eigenvalue, we may expect a negative SRPA eigenvalue with positive norm and a positive eigenvalue with negative norm (a pair of {\em antinormal} solutions, as will be defined below).

We may write the SRPA problem also as an energy-dependent problem of dimension $2M_{ph}\times 2M_{ph}$. 
For a two-body Hamiltonian the SRPA matrix is reduced to
\[ 
\left( 
\begin{array}{cc} 
 A(E ) & B_{11} \\ 
 -B^{\ast}_{11}    & -A^{\ast}( -E) 
\end{array}  
\right) 
\] 
with 
$A(E) = A_{11} + A_{12}(E - A_{22})^{-1} A^{\dagger}_{12}$.  
In the $2p2h$ space we may introduce the representation which diagonalizes $A_{22}$. Then we have 
\begin{equation} 
[A(E)]_{ph,p'h'}  = [A_{11}]_{ph,p'h'} + \sum_{\alpha}[A_{12}]_{ph,\alpha} [A^{\dagger}_{12}]_{\alpha , p'h'} / (E-E_{\alpha})  
\end{equation} 
where $E_{\alpha}$ are the eigenvalues of $A_{22}$.  
This equation proves useful in demonstrating the generation of SRPA instabilities and energetic shifts. 
Let us begin with RPA and 
introduce a single $2p2h$ configuration of high energy--higher than most $ph$ energies. 
Certain matrix elements of $A_{11}$ (including the diagonal) will acquire energy-dependent contributions. 
For $E$ lower than typical $2p2h$ energies, the contributions on the diagonal will all be negative, meaning that, on average, the eigenvalues of $A(\pm E)\pm B$ decrease. 
More than one $2p2h$ configurations will enhance the effect on average, since $2p2h$ energies are higher than $ph$ energies on average.  
The above arguments sketch, albeit incompletely, why lower-lying RPA solutions, including spurious states, appear to shift to lower energies when the space is extended to SRPA~\cite{PaR2009,PaR2010,GaC2009}. 
Now suppose we are working with an operator such that, in RPA,  
$A+B$ is positive semi-definite and $A-B$ is positive-definite (or vice versa). 
After introducing $2p2h$ configurations, $A+B$ need not retain its positive semi-definitness, but could acquire a negative eigenvalue. 

In this connection we point out that certain matrix elements in the SRPA matrix do not affect the HF and RPA problems: In particular, the $3p1h$ and $1p3h$ matrix elements found in the $A_{12}$ matrix. 
This means that we can include randomly huge matrix elements in the $A_{12}$ matrix, thus dramatically altering the SRPA outcome and even deliberately introducing negative eigenvalues, without ever seeing 
any effect in HF and RPA. 
In Ref.~\cite{GGD2012} it is proposed to exploit this apparent freedom to 
better constrain the interaction  
and 
moderate the energetic shifting of resonances.   
The point remains, that zero-energy spurious solutions and always-positive eigenvalues are not guaranteed in SRPA and the reason lies in the formalism itself. 

It can be shown~\cite{Yan1987,AdL1988} that the energy-weighted sum 
associated with any single-particle excitation operator is the same in RPA and SRPA, as long 
as the same HF reference state and the same $ph$ space is used. 
An important outcome of the present work is to reconcile this well-known result with the appearance of 
finite-energy spurious states in SRPA.

\section{The ``fine print": stability condition \label{Sec:Stability}} 

\subsection{Violation of the stability condition~\label{Sec:Violation}} 

A necessary condition for the HF solution to minimize the expectation value of the Hamiltonian in the variational model space consisting of single Slater determinants is that the (first-order) RPA stability matrix be positive semi-definite~\cite{Tho1960,Tho1961}. Satisfying this condition by no means implies that the SRPA stability matrix, namely a supermatrix comprising the RPA stability matrix, will be positive (sem{i-)}definite as well. 
As a result, and as already sketched, it remains perfectly possible to obtain SRPA eigenstates with 
a) imaginary (even complex) eigenvalues or 
b) positive (negative) eigenvalues but with negative (positive) norm (see Eq.~(18) in Ref.~\cite{Tho1961} and remarks in Ref.~\cite{Pap2007}), 
contrary to common assumptions in the literature~\cite{TSA1988,Yan1987}, valid only for RPA. 
We will see numerical realizations of the above cases later on. 

The physical origin of the instability can be grasped with the help of the variational approach by da Providencia~\cite{daP1965}.  
For the reference state to minimize the expectation value of the Hamiltonian, it is required that the 2nd and 3rd term on the right-hand side of Eq. (4) in \cite{daP1965} vanish for all possible $ph$ and $2p2h$ variations. 
Of course, in first-order RPA the $2p2h$ amplitudes are set to zero from the beginning. 
The $2p2h$ space is ignored.  
If the reference state minimizes the expectation value of the Hamiltonian with respect to $ph$ perturbations, then the RPA stability matrix is positive semi-definite, the first-order correction is zero, and the reference state satisfies the HF equations. 
On the other hand, for the reference state to minimize the expectation value of the Hamiltonian in the $ph+2p2h$ space, the stability condition becomes that the SRPA stability matrix be positive semi-definite and the first-order {\em and} the second-order correction vanish. 
This latter condition is simply impossible for a single Slater determinant in all physical cases of interest. 
Therefore, when we build the SRPA matrix using the HF solution as the reference state, we have no grounds for expecting the stability matrix to be positive (semi{-)}definite. 
It should come as no surprise if second-order perturbative corrections lower the ground-state energy with respect to HF.  

Our contentions are not in contradition to the recent derivations of Ref.~\cite{Tse2013}, where positive semi-definitness 
was proven on the condition that a subtraction procedure be 
applied. 
Our derivations for the usual SRPA correspond to the case $\kappa =0$ (no subtraction) in that work. 
The role of high-order correlations in satisfying the stability condition has been pointed out explicitly in Ref.~\cite{ToS2010exp}.  

\subsection{Orthogonality, normalization, completeness, and Thouless' theorem} 

The indefiniteness of the stability matrix affects some of the derivations of Sec.~5 in Ref.~\cite{Tho1961}. 
The proof of orthogonality holds for all eigenvectors with real eigenvalues. 
For such vectors the normalization 
\begin{equation} 
|X_{\nu}|^2 - |Y_{\nu}|^2 = N_{\nu} = \pm 1 
\end{equation} 
(Eq.~(22) in \cite{Tho1961}) is possible. 
It does not follow automatically, however, that positive eigenvalues correspond to vectors of positive norm: 
if in 
\begin{equation} 
(X_{\nu}^{\dagger} Y_{\nu}^{\dagger} ) 
\left( 
\begin{array}{cc} 
  A  &  B  \\ 
  B^{\ast} & A^{\ast}  
\end{array} 
\right) 
\left( 
\begin{array}{c} 
X_{\nu} \\ Y_{\nu}  
\end{array} 
\right) 
= E_{\nu} (|X_{\nu}|^2 - |Y_{\nu}|^2) 
\end{equation} 
(Eq.~(18) in \cite{Tho1961}) 
the matrix is not positive definite, it is possible to find an eigenvector whose norm and eigenvalue have opposite signs~\cite{Pap2007}--it 
becomes, in fact, necessary. 
Let us call such solutions antinormal for convenience. 
In particular, we may introduce the notation 
\begin{eqnarray*} 
&&\mbox{normal solution} \,\, |{n}\rangle :  \,N_{n}E_{n}>0  
\\ 
&&\mbox{antinormal solution} \,\,  |\bar{n}\rangle :  \,N_{\bar{n}}E_{\bar{n}}<0  
\,\, ,  
\end{eqnarray*} 
while Greek letters $\nu$ will denote any kind of eigenvalue.  
It still follows that the norm can only be zero if $E_{\nu}$ is zero. 
It also follows that the set of SRPA solutions is complete, as long as there are no zero-energy (or complex) solutions. 
The unity matrix in the complete $2M\times 2M$ space spanned by the solutions $|\nu \rangle$ can be expanded as: 
\begin{equation} 
\mathbb{I} = \sum_{\nu} N_{\nu} 
\left( 
\begin{array}{c}
X_{\nu} \\ Y_{\nu} 
\end{array} 
\right) 
(X^{\dagger}_{\nu}  -Y^{\dagger}_{\nu})  
\, . 
\label{Eq:unity} 
\end{equation} 
It is straightforward to show that the above expansion acts as the unity matrix on any of the basis arrays. 
The total number of states is even: For each positive-norm solution there will be a negative-norm solution of {\em opposite} (not necessarily {\em negative}) energy. 
 
The variational principle proven in Sec.~5.3 of Ref.~\cite{Tho1961} is modified in the presence of antinormal solutions: then the lower limit is no longer $|E_{\min}|$, 
but a negative number, namely  
$-E^{\bar{n},\max}$, where  
$E^{\bar{n},\max}$ 
is the highest one amongst positive antinormal eigenvalues.

Let us now prove Eq.~(34) of Ref.~\cite{Tho1961} in the general case where antinormal states are present. 
In particular, in this general case  
\begin{equation} 
\sum_{\nu : N_{\nu} = 1} 
|\langle \nu | O | 0 \rangle |^2 E_{\nu} 
= 
\frac{1}{2} 
\langle 0 | [ O,H,O] | 0 \rangle 
\, , 
\label{Eq:ewsrnew} 
\end{equation} 
i.e., the sum spans positive-norm states, of which antinormal states (if any) contribute with a negative sign. 
The energy-weighted sum thus defined is in agreement with the one of Ref.~\cite{AdL1988}. 
Notice that, in general,  
\begin{align*} 
&2 \sum_{\nu : N_{\nu} = 1} 
|\langle \nu | O | 0 \rangle |^2 E_{\nu} 
 = 
2 \sum_{\nu : E_{\nu}>0 } 
|\langle \nu | O | 0 \rangle |^2 N_{\nu}E_{\nu} 
\\ 
&= 
\sum_{\nu} 
|\langle \nu | O | 0 \rangle |^2 N_{\nu}E_{\nu} 
 \neq 
\begin{cases}     
\sum_{\nu : E_{\nu}>0 } 
|\langle \nu | O | 0 \rangle |^2 E_{\nu} 
\\ 
\sum_{n} 
|\langle n | O | 0 \rangle |^2 E_{n} 
\end{cases} 
\, . 
\end{align*} 
One now proceeds exactly as in Eq.~(35) of Ref.~\cite{Tho1961}, but without assuming that $N_{\nu}E_{\nu}=|E_{\nu}|$. 
We obtain: 
\begin{align*} 
&2\sum_{\nu : N_{\nu} = 1} 
|\langle \nu | O | 0 \rangle |^2 E_{\nu} 
=  
\sum_{\nu} 
|\langle \nu | O | 0 \rangle |^2 N_{\nu}E_{\nu} 
\\ 
&=  
\sum_{\nu} N_{\nu}  ( O^{\dagger} -O^T ) 
\left( 
\begin{array}{cc} 
A & B \\ B^{\ast} & A^{\ast} 
\end{array} 
\right) 
\left( 
\begin{array}{c} 
X_{\nu} \\ Y_{\nu} 
\end{array} 
\right) 
(X_{\nu}^{\dagger} -Y_{\nu}^{\dagger})  
\left( 
\begin{array}{c} 
O \\ -O^{\ast} 
\end{array} 
\right) 
\\ 
& = 
( O^{\dagger} -O^T ) 
\left( 
\begin{array}{cc} 
A & B \\ B^{\ast} & A^{\ast} 
\end{array} 
\right) 
\left( 
\begin{array}{c} 
O \\ -O^{\ast} 
\end{array} 
\right) 
\, . 
\end{align*} 
In the last step we have used Eq.~(\ref{Eq:unity}). 
Finally, if the HF solution has been used as the reference state, derivation (37) of Ref.~\cite{Tho1961} follows 
and our statement (\ref{Eq:ewsrnew}) has been proven. 

If the operator $O$ is a single-particle operator, $2p2h$ configurations do not contribute in the evaluation of the right-hand side in 
the above equation. Therefore, as already mentioned, the energy-weighted sum is the same in RPA and SRPA, as long as the same HF reference state is used and the same $ph$ space is considered. 
The same is true for the diagonal approximation to SRPA, whereby the couplings within the $2p2h$ space (i.e., in the $A_{22}$ matrix) are neglected.We shall denote the diagonal approximation as SRPA0.  
 
\subsection{No zero-energy spurious solutions} 

Thouless' theorem on energy-weighted sum rules \cite{Tho1961}, 
i.e., Eq.~(\ref{Eq:ewsrnew}) above, 
has the important consequence that, when the (stable) HF solution is used as the RPA reference state, RPA solutions associated with broken symmetries (translational and rotational invariance, 
in particular)--the so-called {\em spurious} solutions--will appear at zero energy.  
The proof of the above statement can be cast as follows. 
Consider an operator $O$ which commutes with the intrinsic Hamiltonian, 
\begin{equation} 
[ O,H ] = 0 
\, , \label{Eq:commuteOH}
\end{equation} 
i.e., does not generate genuine excitations. We shall call it, quite loosely, a {\em spurious} operator, for convenience. 
Examples of spurious operators include the total, center-of-mass momentum, or the center-of-mass coordinate, $\vec{R}_{\mathrm{CM}}=\frac{1}{A}\sum_{i=1}^{A}\vec{r}_i$, or the 3rd projection of the latter, which is a dipole operator. 
Since the double commutator with the intrinsic Hamiltonian vanishes, 
it follows that 
\begin{equation} 
(O^{\dagger} \,\, O^T ) 
\left( 
\begin{array}{cc} 
  A  &  B  \\ 
  B^{\ast} & A^{\ast} 
\end{array} 
\right) 
\left( 
\begin{array}{c} 
O \\ O^{\ast}  
\end{array} 
\right) 
= 0 
\,\, ,  
\label{Eq:MABM} 
\end{equation} 
which is impossible if the stability matrix is trully positive-definite: the left-hand side would be bound from below by a positive number. 
Since in RPA there cannot be negative eigenvalues,  
the matrix must be positive semi-definite, i.e., have a vanishing eigenvalue, $E_{\mathrm{sp}}=0$. 
The corresponding RPA eigenvector $|sp\rangle $ will be orthogonal to itself. 
The contribution of finite, positive-energy eigenvalues to the energy-weighted sum must vanish, hence no spurious strength can be found at finite energy 
(we shall return to this point in relation to space completeness). 

As already put forth in Ref.~\cite{Yan1987}, 
the above should hold for SRPA too, provided that the SRPA stability matrix were positive-definite. 
As already elaborated, though, the latter expectation is unfounded.  
We stress therefore that 
\begin{quote} 
the existence of spurious solutions at zero energy is a consequence of the positive-definiteness of the RPA stability matrix. 
\end{quote} 
What happens when there are antinormal solutions? 
In that case, the statement that follows from Eqs.~(\ref{Eq:ewsrnew}),~(\ref{Eq:MABM}) is simply that, 
\begin{quote} 
for spurious operators,  
the contribution of the positive-norm antinormal states to the energy-weighted sum (a negative number) will have to cancel out the contribution 
of the other positive-norm states, 
\end{quote} 
so that 
\begin{quote} 
spurious strength need not be concentrated at zero energy. 
\end{quote} 
In our numerical applications we will find, as in previous work~\cite{PaR2009,PaR2010}, that one SRPA state contributes most of the spurious strength, 
and can be considered spurious, even though 
its energy is finite. Each of the other, physical states contributes very little spurious strength by comparison--but the energy-weighted sum of their spurious strengths  
is exactly the opposite of 
the spurious state's contribution. 

\subsection{Completeness issues (bis)} 

Let us now return briefly to first-order RPA, whose stability matrix is positive-definite. 
Formally speaking, the above derivations, also as they appear in the literature~\cite{Tho1961,Row1968a,Yan1987}, 
assume that the space spanned by the RPA eigenvectors is complete. 
At the same time, it has been shown that the presence of a zero-energy solution makes the space one short of complete~\cite{Tho1961}. 
Why should Thouless' theorem still hold?  

It still holds that, if the stability matrix is positive-definite (as in RPA), spurious states must appear at zero energy. 
The reasoning can proceed {\em reductio ad absurdum}: {\em if} we assume a complete space, {\em then} Thouless' theorem holds, resulting in Eq.~(\ref{Eq:MABM}), which is impossible for a trully positive-definite matrix. {\em Therefore}, the space is not complete (and the matrix is positive {\em semi}-definite). 

That the energy-weighted sum associated with a spurious operator shall vanish has not been proven at this point. 
Indeed, we have made use of Eq.~(\ref{Eq:unity}), which presumes completeness. 
One way to proceed is as follows. Consider the sum of the intrinsic Hamiltonian and a small admixture of a term, 
\begin{equation} 
H = H_2 + \epsilon_1 H' 
\end{equation} 
(e.g., $H'$ may be the center-of-mass kinetic energy) 
such that, for a spurious operator $O$ (e.g., a spatial operator acting on the center of mass), 
\[  [O,H] = \epsilon_1 [O,H'] \Rightarrow \langle 0 | [O,H,O] | 0 \rangle  = \epsilon \, . \] 
Here $\epsilon , \epsilon_1$ are infinitesimal quantities. 
Thouless' theorem, for positive-definite RPA matrices, yields 
\[ 
\sum_{\nu : E_{\nu}>0} 
|\langle 0 | O | \nu \rangle |^2 
E_{\nu} = 
\epsilon /2  \, , \, \,  \text{arbitrarily small} .  
\] 
Each term in the sum must vanish in the limit $\epsilon\rightarrow 0^{+}$, i.e., 
must correspond to vanishing energy (spurious) or vanishing matrix element (physical). 
There are certain discontinuities at exactly vanishing $\epsilon$--most obviously, a zero-norm state must appear suddenly. 
At the same time, algebra aside, 
we expect {\em physical} solutions to display continuous behavior. 
The point remains then that, in the limit of an intrinsic Hamiltonian and a spurious operator, excitations of finite energy have no contribution to the RPA energy-weighted sum, 
owe it to the very small energy (spurious) or the very small transition matrix element (physical). 

The RPA energy-weighted sum in the presence of broken symmetries  
has been discussed in Ref.~\cite{StJ2003}, 
with emphasis on rotational symmetry. 

\section{Numerical demonstrations~\label{Sec:Numerical}} 

\subsection{Relevant considerations} 

Let us consider again an $A-$body operator $O$, which commutes with the intrinsic Hamiltonian of the $A$-body system, 
see Eq.~(\ref{Eq:commuteOH}).  
An operator acting on the center of mass of the system would exmplify such a spurious operator, which does not excite the system. 
Of course, for Eq.~(\ref{Eq:commuteOH}) to hold generally, the Hamiltonian must be trully intrinsic. 
The often-used Hamiltonian 
\begin{equation} 
H_1 = \sum_{i=1}^{A} p_i^2/2m + V  
\end{equation} 
(in an obvious notation) is not intrinsic, because the kinetic-energy part, 
$\sum_ip_i^2/2m$, is not intrinsic. 
In special cases, such as $O=\vec{P}_{\mathrm{CM}}$, the commutator in Eq.~(\ref{Eq:commuteOH}) and the double commutator in Eq.~(\ref{Eq:ewsrnew}) still vanish. 
In other cases, such as the widely explored $O=\vec{R}_{\mathrm{CM}}$, they do not. 
Then Eq.~(\ref{Eq:ewsrnew}) does not hold exactly. 
In general, for the commutator to vanish exactly, the center-of-mass kinetic energy must be subtracted from the above Hamiltonian. 
The intrinsic form reads 
\begin{equation} 
H_2 = \sum_{i=1}^{A} p_i^2/2m - (\sum_{i=1}^A \vec{p}_i)^2/2mA + V  
. 
\end{equation} 
(It is assumed that the potential part $V$ depends on intrinsic variables only, which is not true for density-dependent potentials.)  
Use of $H_2$ instead of $H_1$ will produce different results. 
In practice, the difference may be small, but it is present and has shown up in nuclear structure calculations, see, e.g., Ref.~\cite{HPR2011}. 

From the above it follows that numerical validations of our conclusions on spurious states must be based on intrinsic Hamiltonians. 
The Hamiltonian used in this work is therefore of the form $H_2$ with density-independent intrinsic potentials. 
We will present results with two different two-body potentials: 
The Brink-Boeker potential~\cite{BrB1967} 
and the  $V_{\mathrm{UCOM}}$~\cite{RHP2005}, 
namely a unitary transformation of the Argonne V18 two-nucleon potential. 

It goes without saying that the same Hamiltonian is used to solve the Hartree-Fock variational problem and to build the (S)RPA matrix. 
Moreover, the same harmonic-oscillator space is used to solve the Hartree-Fock equations and to build the (S)RPA matrix.

\subsection{Numerical results} 

We consider spherical nuclei. 
Angular momentum and parity $J^{\pi}$are good quantum numbers, therefore we consider each time a given channel $J^{\pi}$, decoupled from the rest. 
The $A$ and $B$ matrices are real, but it should be pointed out that nowhere in Sec.~\ref{Sec:Stability} were they assumed to be real. 
Our conclusions remain of general relevance. 

When we test sum rules, we must take into account the contributions of all (S)RPA eigenstates. 
The full eigenvalue problem must be solved then, i.e., not for the lowest eigenvalues only. 

\subsubsection{Monopole response of $^{16}O$.} 

We use the Brink-Boeker potential within a harmonic-oscillator basis of nine shells and a length parameter of $1.8$fm. 
SRPA and its diagonal approximation are solved for various $2p2h$ energy cutoffs, up to the exhaustion of the present model space, 
i.e., the highest-energy available $2p2h$ configuration at $277$~MeV. 
In this space, the total numbers of $ph$ and $2p2h$ configurations are $M_{ph}=20$ and $M_{2p2h}=9620$. 
RPA is also solved. 
The isoscalar (IS) and isovector (IV) operators considered are 
\begin{align*}
O_{0^+,\mathrm{IS}} &= e \sum_{i=1}^A r_i^2 
\\ 
O_{0^+,\mathrm{IV}} &= e\!\!\!\!  \sum_{ \pi \, \mathrm{(protons)}} \!\!\!\! r_{\pi}^2 
- 
 e \!\!\!\! \sum_{\nu \,\mathrm{(neutrons)}} \!\!\!\! r_{\nu}^2 
\, . 
\end{align*}
In all cases the energy-weighted sums are found exactly the same to several digits, namely $9.073636\times10^3e^2$fm$^4$ for the isoscalar operator and $1.504700\times 10^4e^2$fm$^4$ for the isovector one, validating spectacularly the numerical implementation. 

It has been observed~\cite{PaR2009,PaR2010}, that very collective states (giant resonances) in SRPA have lower energies than in RPA, but approximately the same strength. 
More to the point, it appears that the energy-weighted sum in the giant-resonance region must be lower in SRPA than in RPA. 
There is no contradiction though: The {\em total} energy-weighted sum remains the same, 
but in SRPA a considerable amount is contributed by 
weak and numerous high-energy eigenenstates. 
 
\subsubsection{Dipole response of $^{16}$O} 

The dipole channel $1^-$ includes the spurious operator 
\[ 
O_{\mathrm{sp}} = e 
\sum_{i=1}^Ar_iY_{10}(\hat{r}_i) 
\propto \vec{Z}_{\mathrm{CM}} 
\, . 
\]
In RPA this ``spurious component" to the $1^-$ channel should generate a spurious eigenstate at zero energy. 
In practice (i.e., numerically)  we expect an eigenstate very close to zero, on the positive-real or imaginary axis.  
This we identify as the spurious state. 
In the former case (real value) its transition strength can be calculated (otherwise, it is set to zero). 
We expect practically all strength of the spurious operator to be exhausted by this state. 

Now we consider the electric-dipole operator 
\[ 
O_{E1}^{(1)} = e 
\!\!\!\! \sum_{\pi\mathrm{(protons)}} \!\! \!\! 
r_{\pi}Y_{10}(\hat{r}_{\pi}) 
\, . 
\] 
For intrinsic excitations, this is equivalent to 
the purely intrinsic operator~\cite{BM75}  
\begin{align*}  
O_{E1}^{(2)} = & O_{E1}^{(1)} - \frac{Z}{A} O_{\mathrm{sp}} 
 \\ 
 = & \frac{N}{A} e\!\!\sum_{\pi \mathrm{(protons)}} \!\! 
r_{\pi}Y_{10}(\hat{r}_{\pi}) 
   -\frac{Z}{A}e \!\!\sum_{\nu \mathrm{(neutrons)}} \!\! 
r_{\nu}Y_{10}(\hat{r}_{\nu}) 
\, . 
\end{align*}  
Then 
we expect the spurious $1^-$ state generated by $O_{\mathrm{sp}}$ to contribute practically no transition strength. 
Interesting insights can be gained by looking also at mixed operators, like $O_{E1}^{(1)}$. 

The isoscalar operator 
\[ 
O_{\mathrm{IS}}^{(1)} = e\sum_{i=1}^A r_i^3 Y_{10}(\hat{r}_i) 
\] 
is also mixed. 
It can be written as the sum of the approximately intrinsic operator 
\[ 
O_{\mathrm{IS}}^{(2)} = e\sum_{i=1}^A (r_i^3 - \eta r_i ) Y_{10}(\hat{r}_i) 
 \, , \] 
where $\eta = \frac{5}{3}\langle r^2\rangle$~\cite{VaS1981},  
and the purely spurious operator 
\[ 
\eta  O_{\mathrm{sp}} 
\, .  
\] 
We shall examine the above operators numerically. 
In particular, 
we will gradually increase the $2p2h$ space, by introducing a varying cutoff (a maximal $2p2h$ energy $E_{2p2h,\max}$), 
up to exhaustion of the available $2p2h$ space, 
and examine the behavior of the energy-weighted sums and of the spurious RPA state. 

We study again the $^{16}$O nucleus. 
The Brink-Boeker potential is used in a space of seven shells.  
Then we have $M_{ph}=38$ and, in the full seven-shell space, $M_{2p2h}=11144$.
The small (but sufficient for our purposes) space enables us to solve the full SRPA problem for all eigenvalues and thus
check the sum rules, for large values of $E_{2p2h,\max}$. 
We are also able to solve the SRPA0 problem in the full HF space. 
The length parameter equals $1.8$fm.  

The RPA spurious state is obtained at $i49$~keV (imaginary). 
As we gradually increase $E_{2p2h,\max}$ we can identify an antinormal state as spurious, 
by its strong transition matrix elements of spurious operators (as we will see in the discussion of energy-weighted sums below). 
Its energy becomes more and more negative as we expand the $2p2h$ space.  
Upon exhaustion of the $2p2h$ space the spurious state is found at $-2$~MeV. 
The value can be much more negative in larger harmonic-oscillator spaces: in Ref~\cite{PaR2010}, for example,  
the positive-norm counterpart of the spurious state visible in Fig.~8 is found at about -7.5~MeV. 

\begin{figure} 
\includegraphics[angle=-90,width=\columnwidth]{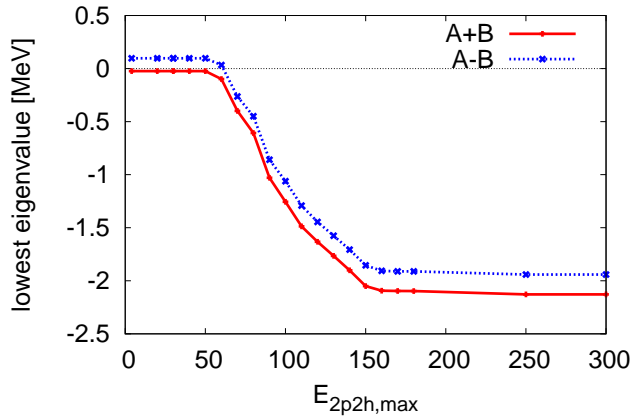}%
\caption{For the $1^-$ channel of $^{16}$O within SRPA 
(see text): 
lowest eigenvalue of the $A\pm B$ matrices 
as a function of the $2p2h$ energy cutoff, up to exhaustion of the $2p2h$ space. 
\label{Fig:Esp}}  
\end{figure}

Since in this case $A$ and $B$ are real, the behavior of the spurious state is controlled by the lowest-energy eigenvalues of the $A\pm B$ matrices, 
see Fig.~\ref{Fig:Esp}. 
In RPA ($E_{2p2h,\max}=0$ or below the first $2p2h$ configuration) 
$A-B$ is positive-definite, while $A+B$ has a negative eigenvalue very close to zero. 
This combination gives RPA an imaginary eigenvalue close to zero. 
For $E_{2p2h,\max}$ larger than approximately 60~MeV, $A-B$ also acquires a negative eigenvalue. 
This combination gives SRPA a pair of antinormal states. 

\begin{figure} 
\includegraphics[width=\columnwidth]{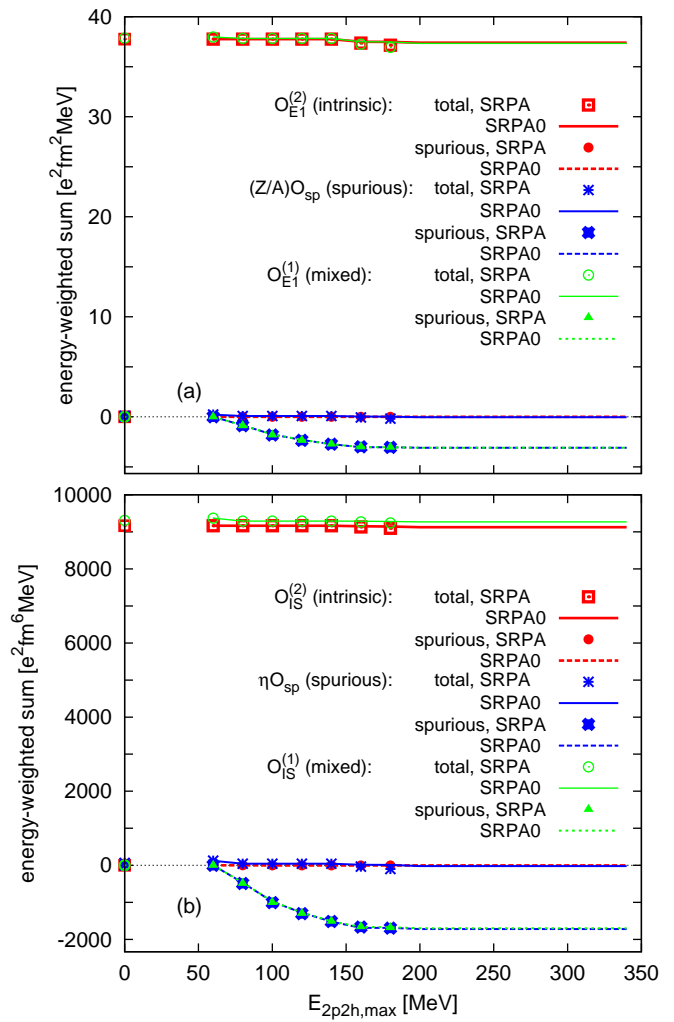}%
\caption{For the $1^-$ channel of $^{16}$O within SRPA(0) 
(see text): 
the energy-weighted sums and spurious-state-only contributions for various dipole operators defined in the text, 
as a function of the $2p2h$ energy cutoff, up to exhaustion of the $2p2h$ space. 
(a) Isovector operators. (b) Isoscalar operators. 
\label{Fig:m1}}  
\end{figure}

Let us now discuss Fig.~\ref{Fig:m1}. 
Here are shown the energy-weighted sums defined by Eq.~(\ref{Eq:ewsrnew}), where the antinormal, spurious state 
contributes a negative quantity. 
In Fig.~\ref{Fig:m1}~(a) the sums are shown for the three isovector operators 
$O_{E1}^{(2)}$ (intrinsic),  
$O_{sp}^{(2)}$ (spurious), 
and  
$O_{E1}^{(1)}$ (mixed). 
The total sums within SRPA and SRPA0 are shown, as well as the contribution of the spurious state only. 
The points corresponding to RPA, namely $E_{2p2h,\max }=0$ in Fig.~\ref{Fig:m1}, were obtained after scaling the residual interaction by 0.999, 
which brings the spurious state to the real value of 18~keV.  
All other results were obtained without any scaling. 

First we observe that the SRPA and SRPA0 results practically coincide, as they should.  
We now look at the SRPA(0) results as a function of $E_{2p2h,\max}$, i.e., as we enlarge the $2p2h$ space.  
We observe readily that the total energy-weighted sums for all operators are the same in SRPA(0) as in RPA, regardless of $E_{2p2h,\max}$. 
The spurious state does not contribute to the sum for the intrinsic operator, while the spurious operator 
gets practically opposite contributions from the physical states (not shown) and from the spurious one, so that the shown total sum is practically zero 
regardless of $E_{2p2h,\max}$. 
We observe, furthermore, that the {\em total} sum for the mixed operator is very close to that for the intrinsic operator, even though the 
contribution of the spurious state changes significantly--and, accordingly, the contribution of the physical states (not shown) changes in the opposite direction.  
Finally, the spurious-state contribution to the mixed-operator sum equals its contribution to the spurious-operator sum ($\frac{Z}{A}O_{\mathrm{sp}}$ is the spurious component of the mixed operator). 

In Fig.~\ref{Fig:m1}~(b), 
the isoscalar operators 
$O_{\mathrm{IS}}^{(1)}$ (mixed), 
$\eta O_{\mathrm{sp}}$   (spurious), 
and $O_{\mathrm{IS}}^{(2)}$ (approximately intrinsic) 
are examined. 
The conclusions are the same as for the isovector operators: 
All energy-weighted sums are practically the same in SRPA(0) as in RPA, regardless of $E_{2p2h,\max}$. 
The intrinsic-operator sum has practically no contribution from the spurious state, 
while the mixed-operator sum receives a negative contribution from the spurious state. 
The sums for the approximately intrinsic and mixed operators are slightly different already in RPA. 
Again, the spurious-state contribution to the mixed-operator sum equals its contribution 
to the sum for the spurious operator $\eta O_{\mathrm{sp}}$. 
 
The above results validate the generalized expression (\ref{Eq:ewsrnew}), i.e., that 
\begin{quote} 
the occurance of SRPA spurious states at genuinely finite energies does not contradict Thouless' theorem.
\end{quote} 
It is important to observe, in this context, that 
\begin{quote} 
in SRPA there are finite spurious admixtures within the physical spectrum.
\end{quote} 
Indeed, the energy-weighted sum for the spurious operator receives contributions from the physical solutions adding up to the same amplitude as the contribution of the finite-energy spurious state. The opposite sign makes the total sum vanish.  
Similarly, the contribution of physical states to the energy-weighted sum for a mixed operator is not the same as for the corresponding intrinsic operator. 
Therefore in SRPA(0), unlike self-consistent RPA, the use of intrinsic operators is of essence.

\begin{figure} 
\includegraphics[width=\columnwidth]{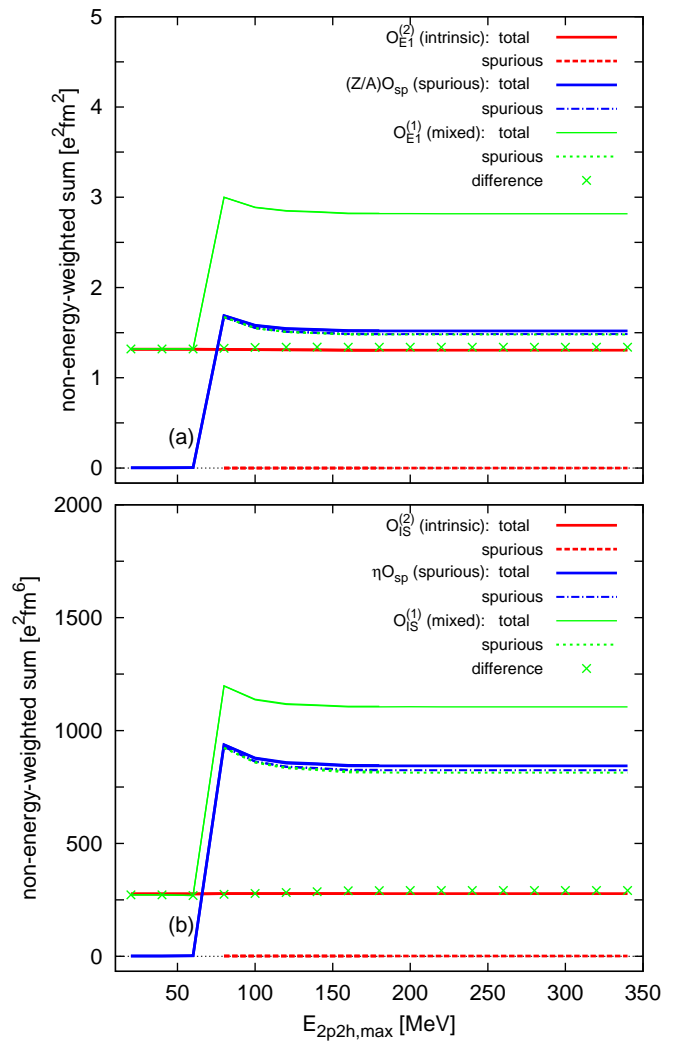}%
\caption{For the $1^-$ channel of $^{16}$O within SRPA(0) 
(see text): 
the non-energy-weighted sums and spurious-state-only contributions for various dipole operators defined in the text, 
as a function of the $2p2h$ energy cutoff, up to exhaustion of the $2p2h$ space. 
(a) Isovector operators. (b) Isoscalar operators. 
Lines between $E_{2p2h,\max}=60$ and $80$~MeV are drawn only to guide the eye. 
The discontinuity is caused by the presence of an imaginary solution for low cutoffs (and in RPA). 
\label{Fig:m0}}  
\end{figure}

Let us take the opportunity to examine also the non-energy weighted sums, 
\[ 
m_0 = \sum_{\nu , N_{\nu}=1} 
|\langle 0 | O^{(1)} | \nu \rangle |^2  
\, .\] 
$m_0$ is expected to 
be (practically) the same in RPA and SRPA(0) for one-body operators~\cite{AdL1988}, provided that the same 
reference state and $ph$ space are used. 
We have verified this to hold, though less precisely than the conservation of the energy-weighted sum~\cite{PaR2010}. 
We now ask what happens in the presence of spurious states and provide some answers through Fig.~\ref{Fig:m0}.  
Here we show how $m_0$ varies with $E_{2p2h,\max}$. 
SRPA and SRPA0 results are both plotted, but the difference is indiscernible. 
A line is drawn between the results for $E_{2p2h,\max}=60$ and $80$~MeV, only to guide the eye. 
The discontinuity is caused by the presence of an imaginary solution for low cutoff values and in RPA. 
In both the isoscalar and isovector cases we observe the following: 
i) $m_0$ for the intrinsic operator receives no contribution from the spurious state; 
ii) most, but not all, of the $m_0$ for the spurious operator comes from the spurious state; 
we see again that there are finite spurious admixtures within the physical spectrum; 
iii) the $m_0$ for the mixed operator has significant contributions from both the spurious state and the physical states; 
but the contribution of the physical states, labelled as ``difference" in the figure,  
is almost equal to the $m_0$ corresponding to the intrinsic operator. 

The above results imply that, if we are interested in the energy region of giant resonances, 
which contribute only moderately to the energy-weighted sum, spurious admixtures in SRPA are not a grave issue. 
The separation of spurious and intrinsic strength can be considered good on average. 
An intrinsic operator should be used though. 

The next application illustrates the more serious problem of instabilities in SRPA.

\subsubsection{Quadrupole response of $^{48}$Ca} 

We now look at the quadrupole response of the nucleus $^{48}$Ca. 
The isoscalar and isovector operators considered here are 
\begin{align*}
O_{2^+,\mathrm{IS}} &= e \sum_{i=1}^A r_i^2 Y_{20} (\hat{r}_i)  
\\ 
O_{2^+,\mathrm{IV}} &=  e\!\!\!\!\!\! \sum_{ \pi \, \mathrm{(protons)}} \!\!\!\! r_{\pi}^2 Y_{20} (\hat{r}_{\pi})  - 
e \!\!\!\! \sum_{\nu \,\mathrm{(neutrons)}} \!\!\!\!\!\!\!\! r_{\nu}^2 Y_{20} (\hat{r}_{\nu})   
\end{align*}
The Brink-Boeker interaction does not have a spin-orbit term. 
For the nucleus $^{48}$Ca we use therefore the UCOM interaction, which generates the necessary spin-orbit splittings. 
$^{48}$Ca has two strong quadrupole states: the giant resonance, and a low-lying one 
generated, in RPA, through the $\nu f_{7/2}\rightarrow \nu f_{5/2}$ configuration. 
We solve the RPA and SRPA within a space of seven harmonic-oscillator shells, where $M_{ph}=82$. 
We look at the behavior of the solutions as a function of the $2p2h$ energy cutoff $E_{2p2h,\max}$, 
up to exhaustion of the seven-shell space at $E_{2p2h,\max}=260$~MeV.  
We then have $M_{2p2h}=90229$. 
The harmonic-oscillator parameter equals 1.8~fm. 

\begin{figure} 
\includegraphics[angle=-90,width=\columnwidth]{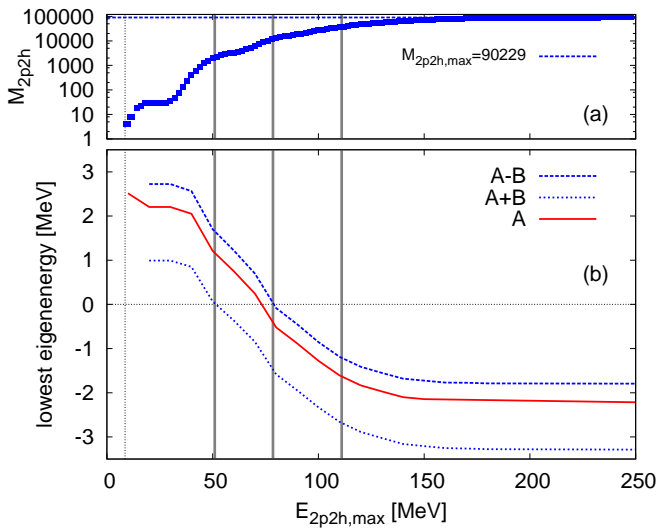}%
\caption{For the $2^+$ channel of $^{48}$Ca within SRPA 
(see text): 
(a) number of $2p2h$ configurations as a function of the cutoff energy $E_{2p2h,max}$, up to 
exhaustion of the $2p2h$ space; 
(b) lowest eigenvalue of the $A\pm B$ and $A$ matrices.  
The dotted vertical line marks the first $2p2h$ configuration. 
Grey vertical lines: see text. 
\label{Fig:camatr}}  
\end{figure}

Fig.~\ref{Fig:camatr}(a) 
shows the number of $2p2h$ configurations available  
as a function of $E_{2p2h,\max}$. 
Fig.~\ref{Fig:camatr}(b) 
shows the lowest eigenvalues of the $A\pm B$ and $A$ matrices. 
The eigenvalues of $A$ are precisely the (S)TDA eigenvalues. 
As expected, the eigenvalue of $A$ remains between the eigenvalues of $A\pm B$. 
They all start off positive, as expected within RPA of this spherical nucleus. 
All RPA solutions are real and positive. 
All STDA solutions are real--but not necessarily positive. 

\begin{figure*} 
\includegraphics[angle=-90,width=0.7\textwidth]{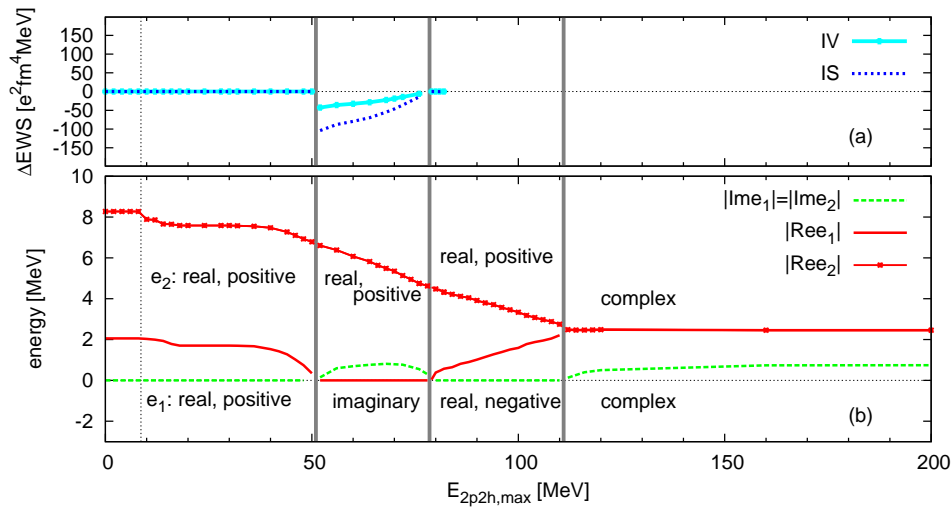}%
\caption{For the $2^+$ channel of $^{48}$Ca within SRPA 
(see text),  
as a function of the $2p2h$ energy cutoff: 
(a) Deviations from the RPA energy-weighted sums.  (b) Behavior of the SRPA eigenvalues, $\pm e_1$ and $\pm e_2$, lying closest to zero. 
In (b) the absolute values of $\Re{e_{1,2}}$ and $\Im{e_{1,2}}$ are plotted, but additional information (e.g., sign of positive-norm eigenenergies, if they are real) is 
given explicitly 
for each $E_{2p2h,\max}$ domain, marked by vertical grey lines. 
The dotted vertical line marks the energy of the first $2p2h$ configuration. 
\label{Fig:caenergy}}  
\end{figure*}

The first two vertical grey lines at about $51$ and $78$ MeV 
mark changes in the sign of the first $A\pm B$ eigenvalues as $E_{2p2h,\max}$ increases: 
First the eigenvalue of $A+B$ becomes negative, then that of $A-B$. 
We expect a pair of imaginary SRPA solutions in this $E_{2p2h\max}$ interval, $51-78$~MeV. 
The energy weighted sums may not be the same as in RPA. 
Beyond the value of $78$~MeV, where each of the $A\pm B$ matrices has acquired one negative eigenvalue, 
we expect a pair of antinormal SRPA solutions. 
Beyond $E_{2p2h,\max}\approx 111~MeV$, marked by the third grey line, we will find a quartet of complex solutions. 
At that point the first positive-norm solution of SRPA and the adjoint of the second positive-norm (and positive-energy) solutions of SRPA have 
acquired the same value, as we will see next, with the help of Fig.~\ref{Fig:caenergy}. 

In the following we denote the lowest positive-norm eigenenergies of SRPA by $e_1$ and $e_2$. 

In Fig.~\ref{Fig:caenergy} are shown 
(a) the deviations of the SRPA energy-weighted sums from the RPA energy-weighted sums for the isoscalar and isovector quadrupole operators 
and 
(b) the behavior of the lowest SRPA eigenvalues, 
as a function of the $2p2h$ energy cutoff, up to exhaustion of the $2p2h$ space. 
 
In RPA the energy-weighted sums for the isoscalar and isovector operators 
amount to, respectively, 
6981 
and 10642~$e^2$fm$^4$MeV. 
In the domain of $e_1$ and $e_2$ real, regardless of norm sign, both energy-weighted sums 
are the same as in RPA within, at worst, a few {\em millionths}--cf. also Fig.~\ref{Fig:caenergy}(a). 
In the domain of imaginary solutions (whose strength is not included in the sums) there are deviations, which is not surprising. 
Note that for $E_{2p2h,\max}>85$~MeV the model space was too large to solve SRPA in full, i.e., for all its eigenstates; 
therefore, sum rules were not calculated beyond that value. 
The behavior of the instability, though, can be tracked. 
In the $E_{2p2h}$ domain where both $A\pm B$ are positive definite (see Fig.~\ref{Fig:camatr}), both $e_{1,2}$ are real and positive. 
The solutions are normal.  
Once $A+B$ acquires one negative eigenvalue, $e_1$ becomes imaginary. 
In the domain where both $A\pm B$ have one negative eigenvalue, $e_1$ corresponds to an antinormal solution. 
Beyond $E_{2p2h}\approx 111$~MeV the two solutions and their adjoint ones coalesce into a quartet with 
same-amplitude complex eigenvalues $\pm e_{1,2}$. 

In Ref.~\cite{PaR2010} the same response was studied within an oscillator space of nine shells. 
In that case SRPA within the full HF $2p2h$ space had a pair of antinormal states. 
The occurance of complex states in the present application may be an artefact of the smaller basis. 
The presence of instabilities remains the persistent fact. 
In going from the 7-shell space to the 9-shell space (both insufficient for convergence), the real part of $e_{1,2}$ shifts from $\pm 2.45$~MeV to $-4.44$ and 0~MeV. 
This is comparable with the energetic shift of the giant quadrupole resonance, namely -2~MeV. 
It has been checked that further increase of the harmonic-oscillator basis 
pushes the antinormal solution to even lower values (more negative).  
Imaginary solutions were furthermore found in the $3^-$ response of $^{16}$O~\cite{PaR2010}. 

It was demonstrated in Ref.~\cite{PaR2010} that ground-state correlations strongly affect the properties of the unstable low-lying solutions, 
but not the higher-lying giant resonances. 
The quadrupole and octupole 
instabilities in the SRPA solutions must, therefore, 
be considered as a genuine shortcoming of SRPA formulated with an uncorrelated reference state. 
HF-SRPA is clearly inappropriate for studying 
these states.

\section{Outlook~\label{Sec:Summary}}

Through a combination of formal arguments and numerical demonstrations, 
the present work establishes that the HF-SRPA method violates the stability condition and 
therefore produces instabilities of no physical origin. 
Spurious strength appears at genuinely finite energy. 
All numerical results are in line with Thouless' theorem. 
It was found, that 
a large amount of spurious transition strength is exhausted by one antinormal state 
and the same energy-weighted amount 
is exhausted by the physical spectrum, so that the total energy-weighted summed strength vanishes. 
The present results further suggest that 
spurious admixtures are not of grave importance in the giant-resonance region, 
although they may be an issue for low-lying transitions--see also Ref.~\cite{PaR2010}. 
On the other hand, the method renders well-known low-lying transitions unstable. 
 
In spite of the problems diagnosed above, 
the HF-SRPA may retain its validity  
when certain conditions are met. 
Existing studies employing exactly solvable models are inconclusive 
as regards the conditions in question~\cite{Toh2007,ShT2003}. 
Apparently, the stronger the residual couplings in the Hamiltonian (and the resulting correlations) are, 
the worse the agreement between SRPA and the exact results is. 
The same holds for RPA, however, and for other extended formalisms. 
Conclusive statements must therefore remain deferred. 
Here we may propose as rather obvious conditions that  
the excitations of interest be described by single-particle operators (e.g., density fluctuations)  
and can be shown to not be sensitive to ground-state correlations, 
and that  
corrected operators be used in the presence of spurious admixtures.  
The above seem to hold for the giant dipole and quadrupole resonances~\cite{PaR2010}, 
but not, for example, for $0\hbar\omega$ transitions 
(cf. the low-lying $2^+$ state of $^{48}$Ca) which are strongly influenced by the depletion of the Fermi sea. 
This issue deserves critical attention in applications of HF-SRPA. 

In the present work we have not employed three-body forces. 
It is unlikely that the addition of three-body forces can eliminate the problematic formal properties of HF-SRPA, 
since the HF solution will always be unstable against $2p2h$ configurations. 
Whether three-body or density-dependent forces moderate the instabilities, by introducing 
many more non-zero terms in the $B$ matrix, remains to be seen in large-scale calculations.  
A restricted $2p2h$ space works to the same effect, as previous applications have shown~\cite{GGC2011,GGD2012}. 
A possibility not considered here is to employ a subtraction procedure~\cite{Tse2013}. 
This method seems attractive in the case of 
interactions which describe excitations well 
already on the level of RPA. 
Then the purpose of SRPA is solely to account for fragmentation. 

For a wider range of applicability, an improved SRPA formalism is necessary. 
In principle, a robust formulation of second-order RPA requires that the correlated reference state 
be the solution of an extended variational problem beyond HF.

\begin{acknowledgments}
I wish to thank Nguyen Van Giai and Marcella Grasso for critical discussions which 
triggered this work 
and Peter Schuck for useful remarks. 
The present work 
was supported by the Rare Isotope Science Project of the Institute for Basic Science funded by the Ministry of Science, ICT and Future Planning and the National Research Foundation of Korea (2013M7A1A1075766). 
\end{acknowledgments}


%

\end{document}